# Dynamic PL&T using Two Reference Nodes Equipped with Steered Directional Antenna for Significant PL&T Accuracy

Niraj Shakhakarmi
Dept. of Electrical and Computer Engineering
Prairie View A&M University
(Texas A&M University System)
Prairie View, Texas 77446, USA

Dhadesugoor R. Vaman
Dept. of Electrical and Computer Engineering
Prairie View A&M University
(Texas A&M University System)
Prairie View, Texas 77446, USA

*Abstract*— **Dynamic Position Location and Tracking (PL&T) is proposed deploying the integrated approach of zone finding and triangulation using two friendly nodes equipped with Steered Directional Antenna (DA) in Mobile Ad hoc Networks (MANET). This approach allows the system to use only two references instead of a typical 3 references for a straight triangulation. Moreover, the performance of the proposed algorithm with references using directional antennas shows significant improvement over triangulation using references with Omni-directional antennas as the beam power is concentrated. However, dynamic switching of reference nodes is frequently required as the target moves outside the predicted zone. This paper presents a better tracking accuracy in using proposed dynamic PL&T as compared to other PL&T techniques. The multipath fading is also addressed with the use of KV transform coding technique which uses forward error correction and sample interleaving achieves greater than 90% tracking accuracy with BERs of $10^{-6}$ or better.**

## I. Introduction

Position, Location and Tracking (PL&T) algorithms based on triangulation require exchange of Time of Departure (ToD) and Time of Arrival (ToA) of packets to generate the range and with three reference nodes which have accurate two dimension (2D) PL&T location, it is possible to track a target node very accurately as long as multi-path fading interference is handled which is very severe indoors [1]. Using four reference nodes, it is possible to achieve 3D PL&T, however the references have to be above the ground (at least one node) at least 100 ft. Usually, the nodes can be placed in Unmanned Aerial Vehicles (UAV) and used as references. Extensive work has been done in Future Warrior Soldier program by Boeing Corporation using Raytheon Micro-lights and more recently Joint Tactical Radios (JTRS). Also, using Internet Protocol (IP) based triangulation for 2D and 3D PL&T tracking using embedded Real Time Predictive Trajectory Polynomial demonstrated tracking accuracies of less than a meter [2]. However, severe multi-path handling in indoor environment is still a serious issue.

One of the critical issues in PL&T tracking based on triangulation is the cumulative error in the PL&T as nodes are continuously tracked at different locations, even though the tracking accuracy at each location is 1 meter. Thus, there is a need for re-initialization of each node after repeated tracking at different locations. This can only be achieved by sending the node to a known location for recalibration. Also, the reference nodes can also move and as they move, their locations can become inaccurate. Thus, there is a need to make them targets with other accurate references to allow them to achieve accurate positions using the PL&T triangulation. Therefore, in triangulation based tracking, there is a need to switch each node as target and as reference which requires dynamic reference module above the PL&T triangulation. The complexity increases and therefore, power consumption in each node increases.

The use of triangulation for PL&T is becoming increasingly important as many countries have capabilities to destroy the satellite, thereby denying Global Positioning Satellite (GPS) based tracking. Also, GPS is not very accurate near the buildings and inside the buildings. The triangulation for PL&T allows MANET to maintain strict friendly nodes for multi-hop connectivity. It is possible to design a system where ToD and ToA are encrypted and used only in strict friendly nodes. However, the issue of multi-path interference needs to be addressed without which one cannot design PL&T system based on triangulation.

## II. Background And Prior Research

The forward movement based prediction considers zone prediction within a constrained forward movement of a target node [3]. It uses single reference node, limited random movement and does not consider sharp turns or obstacles which can be addressed by the optimal zone forming that considers any random movement. In addition, the received signal strength based prediction has better tracking accuracy over the triangulation method as long as multi-path fading is small and longer averaging availability [4]. The signal is spread in the lack of zone and errors are accumulated in the computation of both the distance and the angle and has limited non-random trajectory. This can be resolved by Zone finding and adaptive beam forming.

On the other hand, multi hop based prediction deploy the position locations of nodes estimation using multiple levels of reference nodes which increases the cumulative errors in multi hop measurements and has no beam adaptation used [5]. Thus, significantly reduces accuracy of tracking. This is over-come by zone forming with dynamic switching of reference nodes and dynamic ranging which provides improved accuracy in the location tracking. Furthermore, directional lines intersection based prediction localization of nodes use point of intersection of highly directional beams for low speed mobile anchor nodes [6]. This has higher overhead in scanning and does not address random trajectory. This can be addressed by zone forming and adaptive beam forming which reduces the scanning overhead in the random trajectory.

## III. DYNAMIC PL&T WITH TWO REFERENCE NODES EQUIPPED WITH DIRECTIONAL ANTENNAS

For real time PL&T operation, the use of directional antennas provide methodology of sweeping the antenna and finding a zone prior to the execution of PL&T algorithm for finding the coordinates of the target. This improves the accuracy of PL&T algorithm. The target node is tracked by two nearest friendly neighbors by focusing their steered directional beams over the target node. The dynamic tracking zone is formed using the previous two locations of the target with two friendly nodes and mapping time difference of ToA and ToD packet estimation into corresponding location information. Later directional beams of two friendly neighbors are converged over the tracking zone and dynamically updated till the target is inside the dynamic range of tracking nodes. These friendly neighbors can be dynamically changed as per requirement, in out of range scenario, while tracking the target node. The proposed dynamic PL&T algorithm operates as follows:

- When a target node is detected in the neighborhood range of a node then each target node is tracked by two nearest friendly neighbor nodes $A_j$ and $C_k$ by concentrating their steerable directional beam spreads over the target $B_i$. The fresh initial position of the target node $B_i$ is localized by mapping the time difference between ToA packets and ToD packets into radial distances $d_{Aj,}$ $d_{Ck}$, and Angle of Arrival (AoA) into the direction. This is simultaneously done by both tracking nodes $A_j$ and $C_k$ using Directional Wait to Send (DWTS) based Directional Media Access Control (DMAC) over target node for its initial position localization. It is possible to consider verification of the position with a known wait location.

- After localizing initial position of the target node Bi by tracking nodes Aj and Ck, a tracking zone is developed based on knowing the latest two position information of the target ($B_i$ and $B_{i+1}$). The following procedure is strictly deployed to develop the robust tracking zone.

  ✓ Location initialization of target $\{B_i, B_{i+1}\}$: Packets are exchanged between $A_j$ and the target; and between $C_k$ and the target. Using directional beams, distances $d_{Aj}$ and $d_{Ck}$ are derived from the time difference of ToA and ToD of each packet in an ensemble (Fig. 1). The distances are stretched longer and they form two sides of a triangle. Then, the average distance $d_{average} = (d_{Aj} + d_{Ck})/2$ is marked on the line joining the latest two position of the target $B_i$ and $B_{i+1}$. The base of the triangle is drawn through the marked point along the line parallel to line joining the two reference nodes $A_j$ and $C_k$. This triangulation represents both the position of target node $B_i$ and tracking nodes $A_j$ and $C_k$, to predict the future position of the target.

  ✓ An inscribed circle is sketched inside the triangle at the point of intersection of any two angle bisectors and radius '$r_m$' as perpendicular distance over all sides of the triangle. The radius can be computed by:

$$r_m = \sqrt{(s- d_1)(s- d_2)(s- d_3)/s} \tag{1}$$

  where, $s=(d_1+d_2+d_3)/2$ and $d_1,d_2,d_3$ are sides of a triangle.

  The equation of the inscribed circle is given by:

$$(x - x_m)^2 + (y - y_m)^2 = r_m^2 \tag{2}$$

  ✓ A line sketched through the latest two positions of the target $B_i$ $(x_{i-1},y_{i-1})$ and $B_{i+1}$ $(x_i,y_i)$ is $y = kx+ c$, where k is the slope and c is the y-intercept which are calculated by:

$$k= (y_i-y_{i-1}) / (x_i-x_{i-1}) \tag{3}$$

$$c = (x_i y_{i-1} - x_{i-1} y_i)/(x_i - x_{i-1}) \tag{4}$$

  ✓ Another circle is also drawn such that its radius '$r_n$' is the half of the distance between the latest two position of the target $B_i$ and $B_{i+1}$ in the same directional movement of the target. The equation for the second circle is given by:

$$(x - x_n)^2 + (y - y_n)^2 = r_n^2 \tag{5}$$

  ✓ The point of intersection P $(x_p,y_p)$ of the line joining the target $B_i$ and $B_{i+1}$ and the line joining intersected circle of radii $r_m$ and $r_n$ is the novel point defined as:

$$x_p = \{(x_m+x_n)\,(x_m-x_n) + 2*(r_n^2 - r_m^2)\}/2*(x_m-x_n)\,) \qquad (6)$$

$$y_p = \{(r_m^2 - r_n^2 + x_n^2 - x_m^2 + y_n^2 - y_m^2) - 2*(x_n - x_m)x_p\}/2*(y_n - y_m) \qquad (7)$$

The robust tracking zone is developed by drawing a new circle with diameter $d_o = r_m$ over the line joining the target $B_i$ and $B_{i+1}$ at the novel point where the future position of the target $B_{i+n}$ is localized and tracked inside the intersection of the two beams from the two reference nodes. Figure 1 illustrates the determination of tracking-zone deploying two friendly nodes as references.

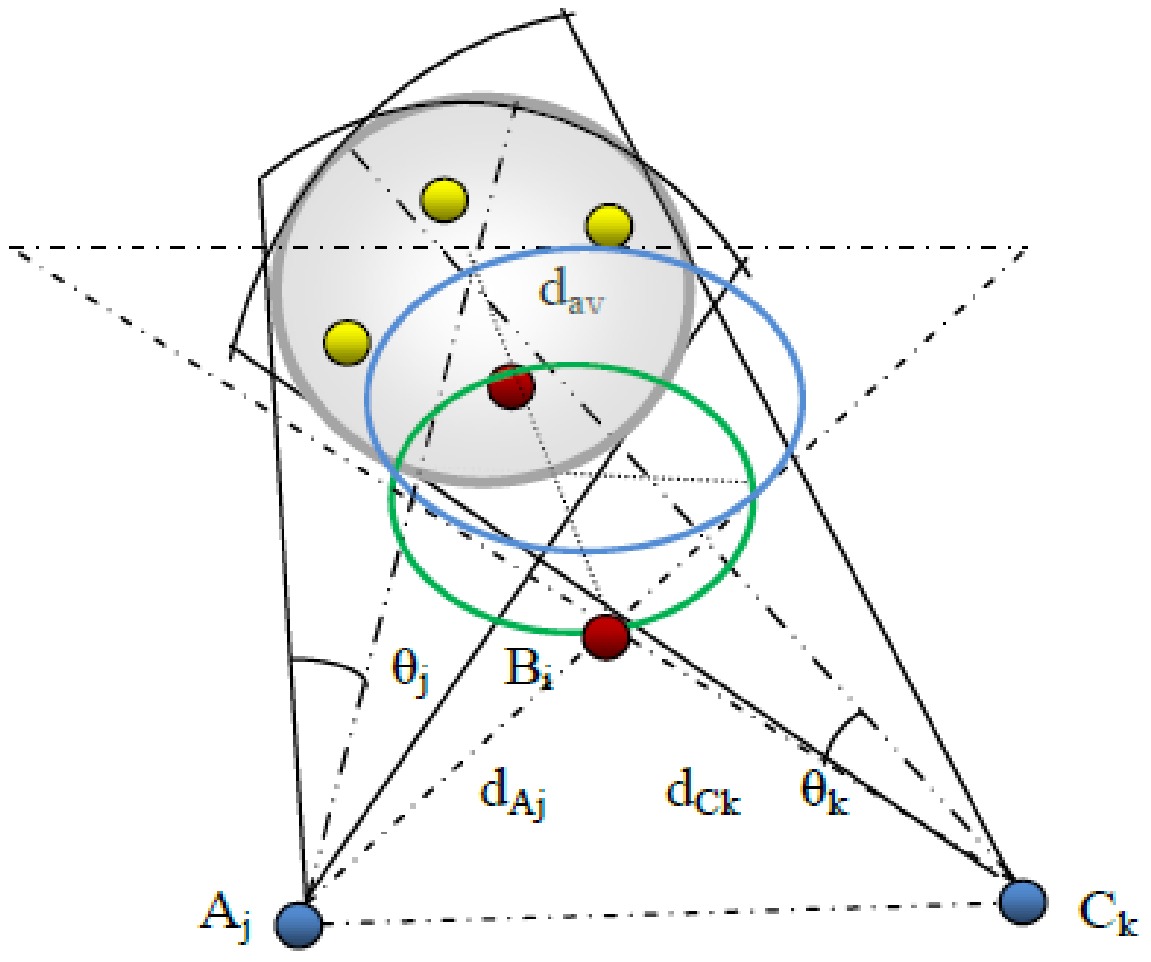

Figure 1. Tracking-Zone deploying by tracking nodes

Once the tracking zone is determined, the two reference nodes and the target node exchange packets which contain the ToD and ToA of each packet. The average of (ToA – ToD) provides the range between the target and any one reference node. Thus, two ranges are determined between the reference nodes and the target. We also assume at an instance of time, the PL&T location of the reference nodes are known and therefore the distance between the two references. By finding the ranges, we also know the distance of the target from each reference. Then a simple triangulation will determine the PL&T of the target node with respect to the reference node. By rotation of the axis, it is possible to determine the PL&T of the target with a particular reference node. If that reference node happens to have a known GPS data, the relative PL&T of the target can be converted to GPS value in 2D (which is the only system considered in this paper).

The beam widths adapted by reference nodes $A_j$ and $C_k$ at distance $D_j$ and $D_k$ from the centre of zone having the zone radius $r = r_m$ are $\Theta_{Aj}$ and $\Theta_{Ck}$ defined by equations:

$$\Theta_{Aj} = 2\arcsin(r/D_j) \qquad (8)$$

$$\Theta_{Ck} = 2\arcsin(r/D_k) \qquad (9)$$

The Euclidean distances and angle of arrivals of reference nodes about the target node are computed as $D_{AB}$, $D_{CB}$ and □$_{AB}$, □$_{CB}$ in equations 10-13.

$$D_{AB} = c*\sum_{i=1}^{n}(ToA_i^{BA} - ToD_i^{AB})/n \qquad (10)$$

$$D_{CB} = c*\sum_{i=1}^{n}(ToA_i^{BC} - ToD_i^{CB})/n \qquad (11)$$

$$\theta_{AB} = \sum_{i=1}^{n}(\theta_i^{AB})/n \qquad (12)$$

$$\theta_{CB} = \sum_{i=1}^{n}(\theta_i^{CB})/n \qquad (13)$$

The location of target node $B(x_{AB}, y_{AB})$ from node A is ($x_{AB} = D_{AB}$ Cos □$_{AB}$, $y_{AB} = D_{AB}$ Cos □$_{AB}$) and $B(x_{CB}, y_{CB})$ from node A is ($x_{CB} = D_{CB}$ Cos □$_{CB}$, $y_{CB} = D_{CB}$ Cos □$_{CB}$) respectively. The position trajectory of a target node A moving with velocity 'v' at time 'Δt' is estimated by node B and node C, are $P_{AB}$ and $P_{CB}$ as follows in equations 14 and 15.

$$P_{AB} = v_{target}(x_{AB} + y_{AB})\Delta t = v_{target} * D_{AB}(\cos\theta_{AB} + *\sin\theta_{AB}) * \Delta t \qquad (14)$$

$$P_{CB} = v_{target}(x_{CB} + y_{CB})\Delta t = v_{target} * D_{CB}(\cos\theta_{CB} + *\sin\theta_{CB}) * \Delta t \qquad (15)$$

Dynamic switching of reference nodes in PL&T is the assignment of new friendly reference nodes as per requirement when the target node goes out of range from the current reference nodes during localization and tracking. Dynamic switching overhead is increased when the velocity of target is comparatively higher than that of reference nodes.

Dynamic Ranging is the process of changing the directional communications range varying the beam width depending upon the target's location. The narrower beam provides the highly concentrated beam for a farther target with the higher range as well as higher directional gain and vice-versa. This improves the data rate, signal quality and tracking accuracy. When the distance d of centre of zone from reference node and the radius of zone r, are determined then the directional beam width $B_w$ is determined as:

$$B_w = 2*\mathrm{ArcSin}\,(r/d) \qquad (16)$$

In addition, the Beam width of adaptive directional antenna is expressed as;

$$B_w = \eta * P_t / R^2 \qquad (17)$$

where, R is the range, $\eta$ is the radiation efficiency and $P_t$ is the transmitted beam energy.

The directional range is inversely proportional to the size of zone and beam width from equations 17-18 which is represented in Figure 2.

$$R=\sqrt{\eta * P_t} * \left(2 * \text{Arcsin}\left(\frac{r}{d}\right)\right)^{-1/2} \tag{18}$$

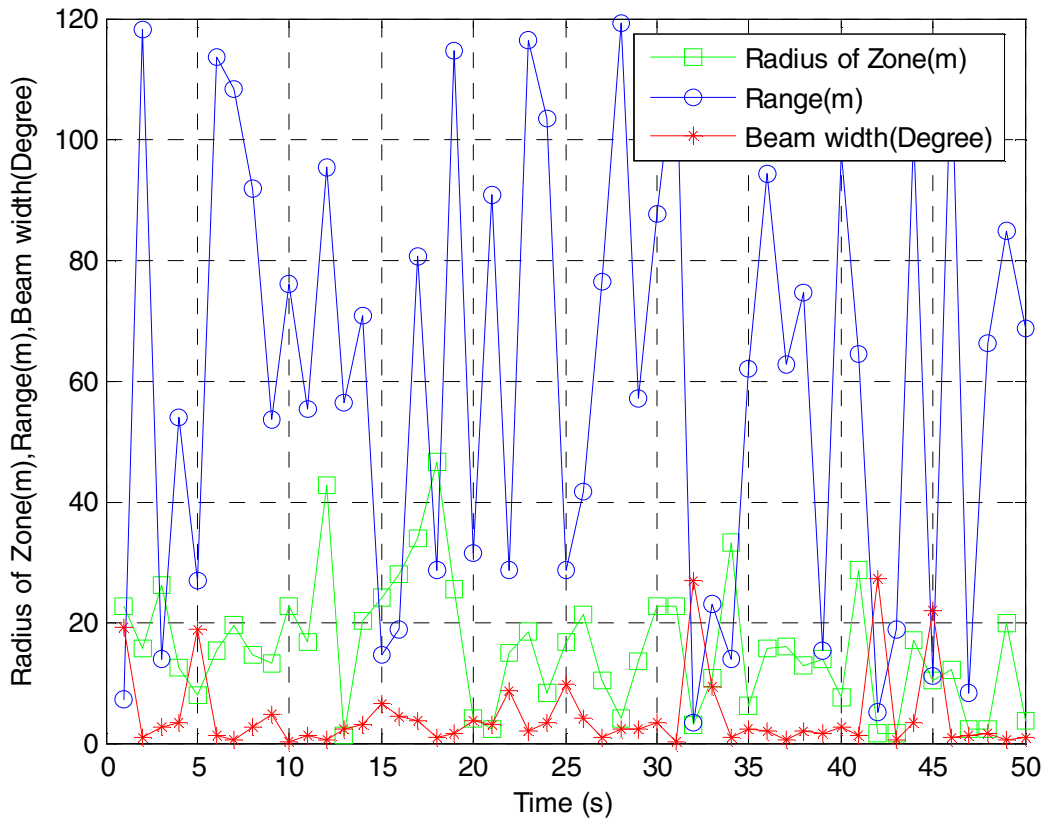

Figure 2. Illustration of Dynamic Ranging

The states of target and reference nodes can be modeled as time dependent systems [7]. Initially, $A_j^{th}$ and $C_k^{th}$ tracking nodes are in searching 'S' mode for target $B_i^{th}$ in the space. In other words, these two nodes are broadcasting the message, and looking for target in their power range. The 'S' mode is replaced by 'T' as $A_j^{th}$ and $C_k^{th}$ nodes keep tracking target $B_i^{th}$ till it is found inside the range. When the target $B_i^{th}$ is in either 'S' or 'T' mode, the internal clock is coupled with mode by $\Delta t$. If the internal clock coupled at mode 'S' marks a time larger than $\tau$, the state of the agent is changed to reset 'R' mode of the clock. If the mode 'R' keeps the clock for a time larger than $R_\tau$, the agent moves back to the 'S' mode. The transition from 'S' to 'T', and from 'R' to 'S' are finite processes.

## IV. Comparison Of Proposed Dynamic PL&T With Other PL&T Schemes

In a random trajectory, the average errors are found 1.25 m in Dynamic PL&T, 5.51 m in location prediction by directional communications (LPDC), 7.27 m in gain direction of directional line intersection (GDDI), 7.11 m in angle of arrival (AoA) and 11.47 m in received signal strength (RSS) method considering 10 different random experiments as shown in Figure 3 and Figure 4. This concurs that the RSS method has higher error as the RSS is highly fluctuated under fading channels. In addition, AoA and GDDI method have similar error range as both deploy AoA and RSS. Furthermore, LPDC has forward zone constrained which is improved by DPL&T forming zone for random trajectory to reduce error significantly. The average location errors are improved by 72 %, 80 %, 83%, 89 % in Dynamic PL&T as compared to location prediction by directional communications (LPDC), gain direction of directional line intersection (GDDI), angle of arrival (AoA) and received signal strength (RSS) method.

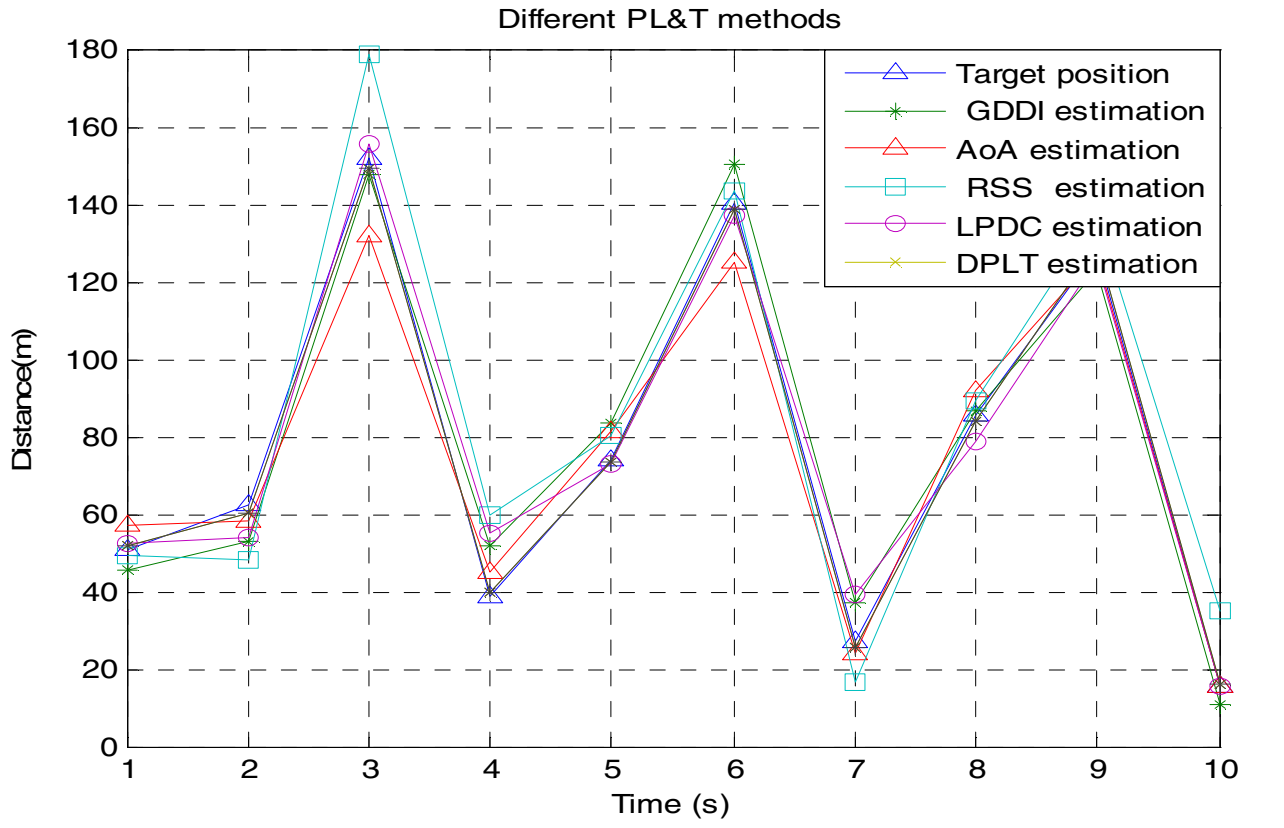

Figure 3. Comparison of Different PL&T Methods

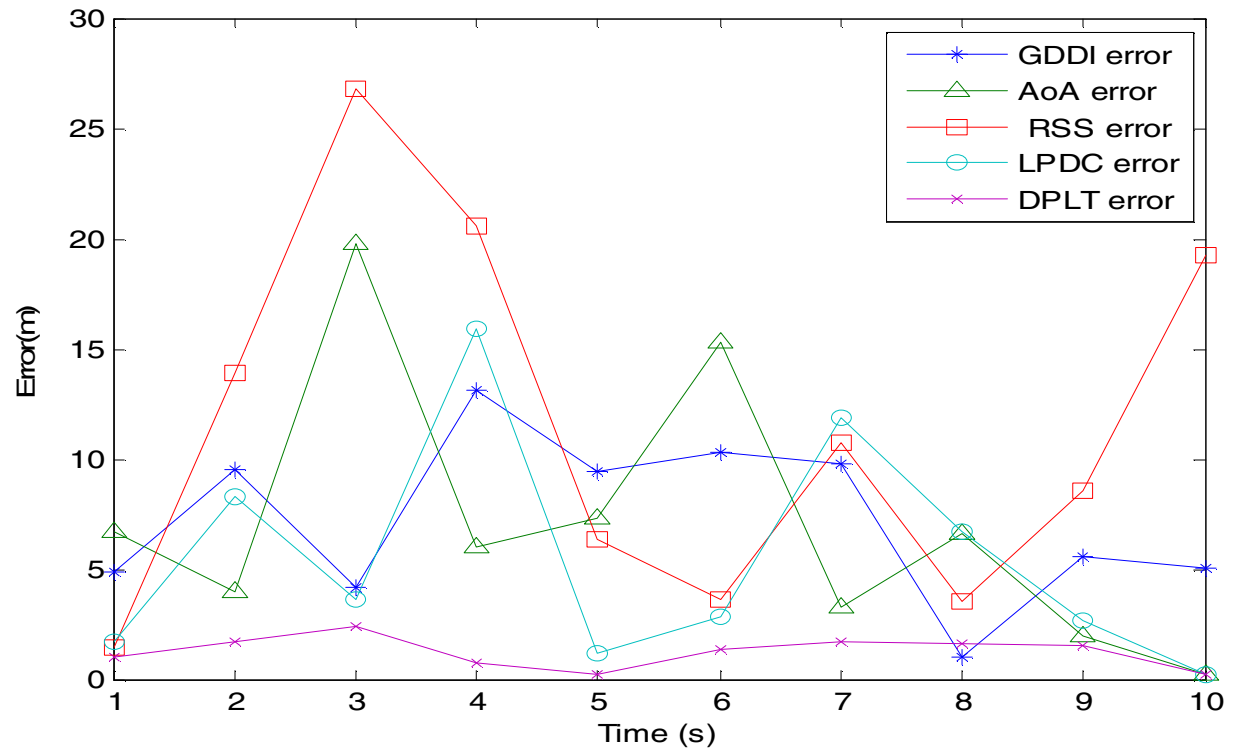

Figure 4. Errors in Different PL&T Methods

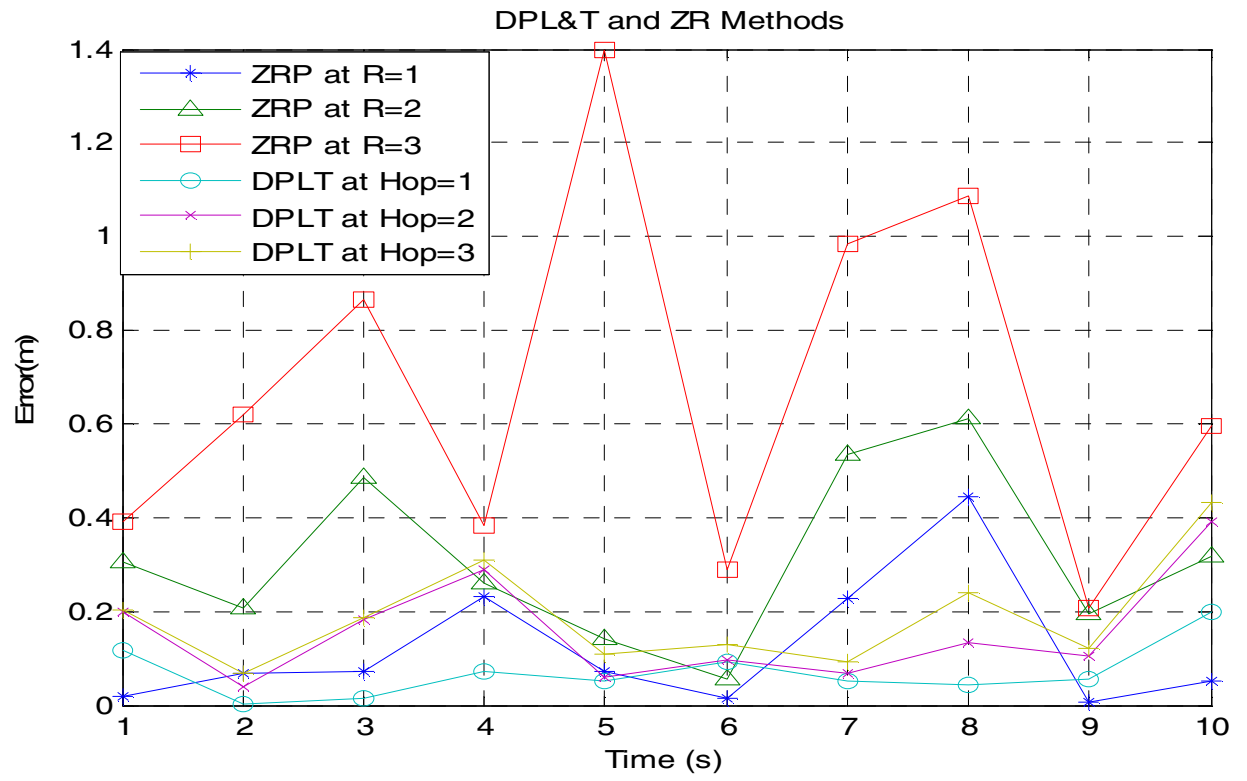

Figure 5. Comparison of DPL&T and Zonal Routing

While comparing PL&T performance of Zone Routing (ZR) with Dynamic PL&T, it is found that the PL&T error is lower in Dynamic PL&T as compared to Zonal routing method as shown in Figure 5. The major reason is that Zonal routing deploys the range-free method in which each node is localized based on number of hop count within the specific zone radius [8]. The zone radii are based on hop radii R=1, 2, 3 considering directional steering for locating nodes. On the other hand, dynamic PL&T develop the narrow zone and adaptive beam forming for locating node and iteratively keep for number of hops=1,2,3. From simulation, DPLT has significantly lower error in single hop based location and similar results in two hops and three hops using DPLT as compared to ZR using single hop radius.

## V. SIMULATION & PERFORMANCE EVALUATION

In our simulation, we implemented dynamic PL&T operation using MANET clusters consisting of 60 nodes in a 500 x 500 sq. m cluster area where each node can be a reference to a given target. The channel Frequency is 2.54 GHz, radiation efficiency is 0.82, antenna size is 1 m with 5 elements at equal spacing, adaptive beam forming, transmitted power is 40dbm and maximum range is 300 m to compute received power by target node from both Reference nodes.

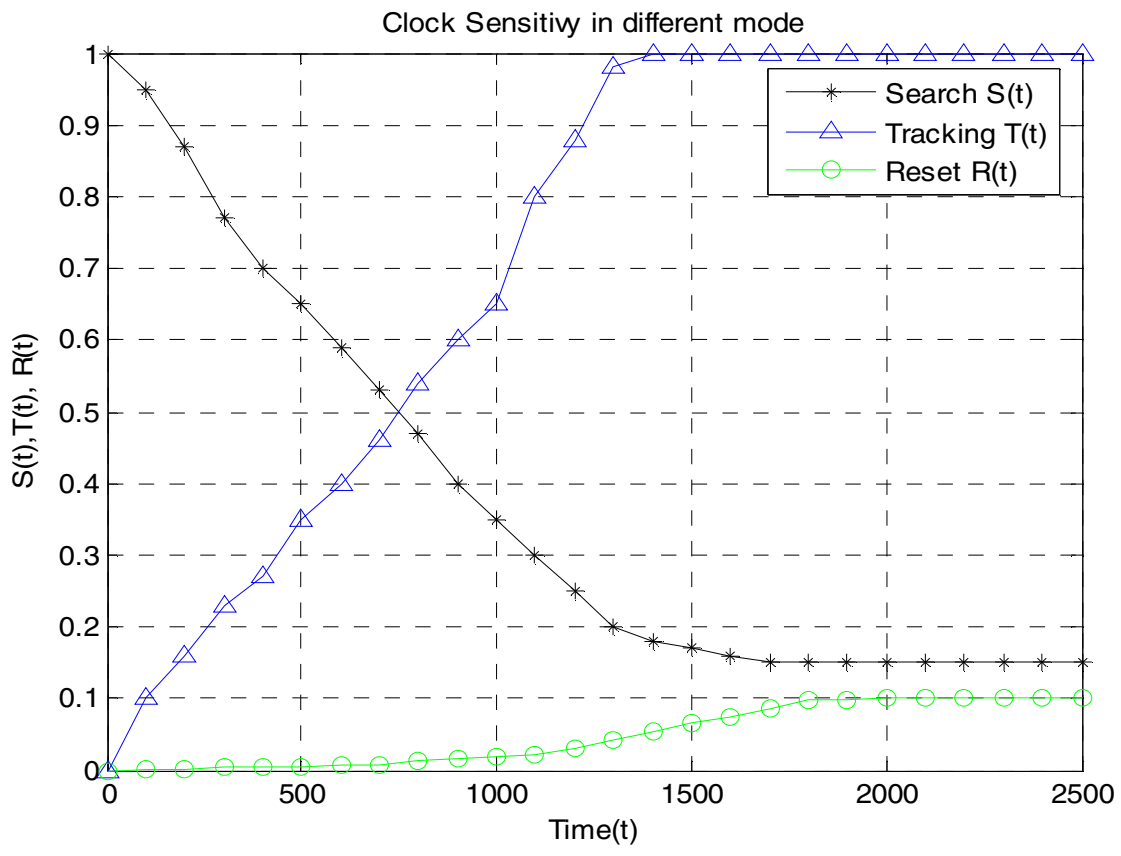

Figure 6. Clock Sensitivity Illustration

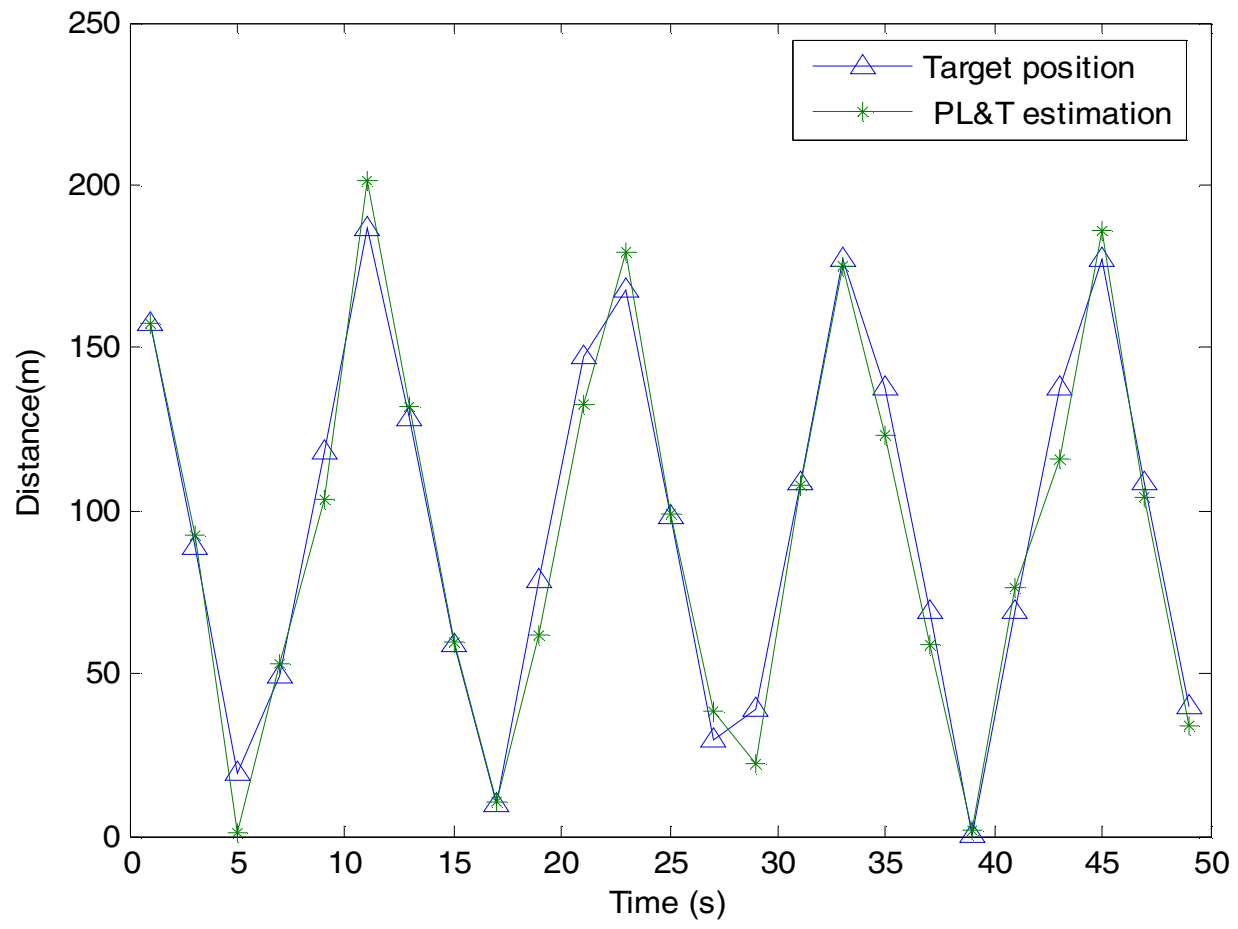

Figure 7. PL&T Estimation

Regarding clock sensitivity, a node coupled with internal clock 't' is in searching mode, S(t) for scanning a target. If the time in S(t) mode exceeds the threshold time ($\tau$) = 1500 ms, then it moves into reset mode, R(t). On the other hand, when two nodes find the target then they switch into tracking mode, T(t) and stay in T(t) mode by changing their position dynamically until the target is inside the range. The point of mode equilibrium exists at t = 750ms at which the PL& T mode switching can occur from S(t) to T(t) and vice versa. In other words, tracking nodes could switch from S(t) mode to T(t) on the target detection and T(t) mode to S(t) on missing target as shown in Figure 6.

PL&T performance is evaluated in terms of distance over time such that target is moving in random direction at different speed in the range of 10m/s and 40 m/s. The speed is lower only when the target makes the sharp turn along its path and higher along straight path. The tracking nodes search the target in their range and form tracking zone over the target by the intersection of the directional beams. The PL&T estimation is performed and the tracking zone is dynamically formed depending upon the target's position. Figure 10 shows the average PL&T error as a function of speed of the node movement. This illustrates the comparison between "true target position" and "PL&T estimated position" for a range of distance and time instances. The average PL&T error is 4.16 m in Figure 6 scenario.

The average error of PL&T is demonstrated at different speed with adaptive beam forming as well as Omni-directional mode in Figure 7. It is found that the average error goes on increasing with increasing speed of target because of multipath fading and Doppler spread. The average error percentage is found highest in the lack of robust coverage and scalable tracking zone in single node using adaptive beam forming DA, then triangulation based PL&T with two nodes having OA, three nodes having OA and the lowest in two nodes having directional antenna for adaptive beam forming based Dynamic PL&T. The minimum average error is 0.05 m at 5 m/s and maximum average error is 1.35 m at 50 m/s in proposed two nodes based PL&T using directional antenna which proves that the proposed PL&T is very efficient as compared to PL&T scheme deployed with single node using DA, two nodes using OA and three nodes using OA.

PL&T simulation shows that the average broadcasting time increases with the increasing probability of changing directions by target because the target always move away in other random direction and it consumes more time to allow the data to be successfully received by the target. In addition, the beam width of reference nodes plays a vital role such that narrower beam width provides the long range which significantly increases the speed of broadcasting packets and consumes lower broadcasting time. This is sustained as the average broadcasting time is found the most efficient in 15 degree beam width than 30 degree beam width, 45 degree beam width, 60 degree beam width, 90 degree beam width and Omni-directional, even the probability of turning of target is increased. Figure 8 shows the Average Broadcasting Time over probability of changing direction by the target.

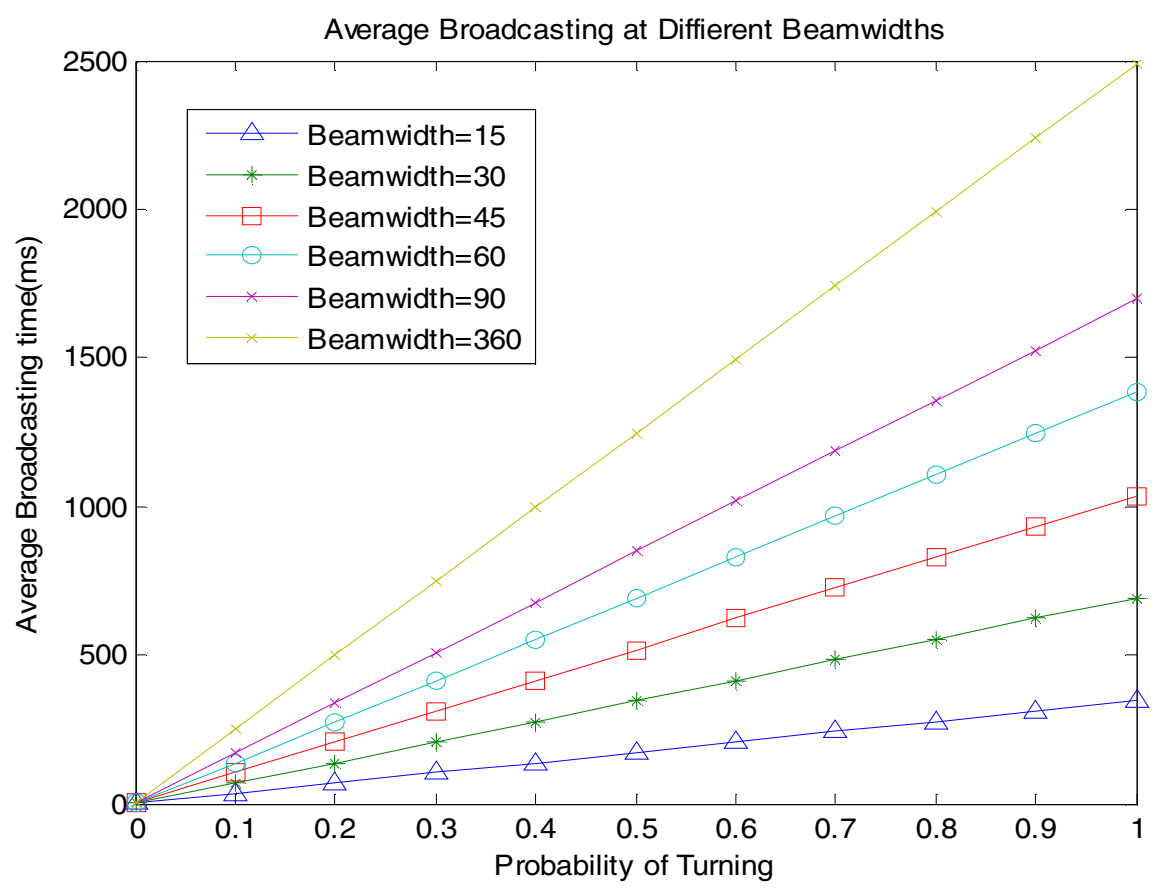

Figure 8. Average Broadcasting Time over probability of changing direction by target

Transmitter receiver pair gain of PL&T is analyzed over time with random beam width and rotation of nodes as shown in Figure 9. The rotation angle is dominant in gain when it is higher than the beam width and is small otherwise. The gain varies between 4.5db and 5.7 db over 100 seconds where lower beam width or rotation angle provides higher gain.

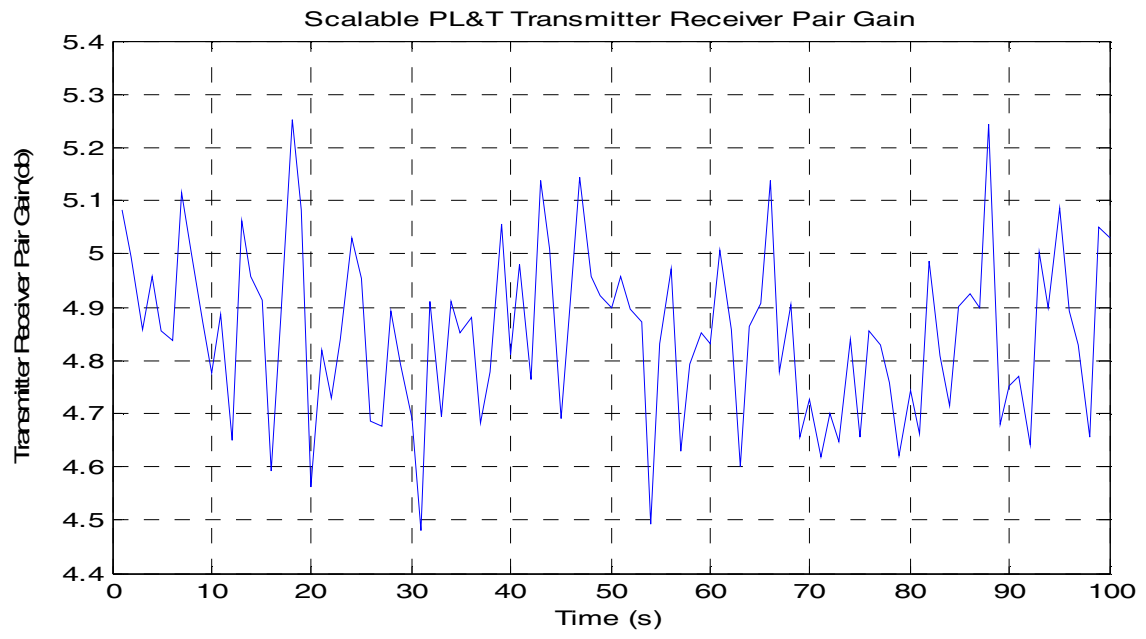

Figure 9. Transmitter Receiver Gain

## VI. Tradeoff Between Beamwith And Tracking Accuracy

When the tracking zone is wider, the target can be exactly localized and tracked inside the same zone for higher number of trajectory trials until target does not move away. This increases the tracking efficiency but reduces tracking accuracy as the beam is spread over large coverage. On the contrary, when the zone is smaller, the target falls outside the existing zone and need to update the zone. This reduces the tracking performance as the computational time overhead is accumulated when the zone is frequently updated. In addition, when the zone is narrow the beam width is also narrow and it increases the tracking accuracy as well as data transmission rate. This is the tradeoff between the zone size and tracking accuracy which can be manipulated as the converse relation between the beam width and zone updating time overhead. The tradeoff caused by beam width, zone updating overhead and PL&T estimation error can be empirically related together as shown below from the simulation.

$$B_W=10/T_{Zonal}=P_{error}/0.025 \quad (19)$$

PL&T Performance including estimated error and zonal overhead with respect to Beam width. With increasing beam width, zonal overhead for updating zone decreased but the PL&T estimation error increased. The estimated error has some decreasing jitters because of line of sight which decreases error even in higher beam width. From the simulation results, the critical beam width for Dynamic PL&T is found 10 degree above which the zonal overhead decreases and estimation error increases significantly. The optimum beam width for Dynamic PL&T deploying narrow zone forming and triangulation using two reference nodes, is 10 degree to optimize the estimation error and zonal overhead.

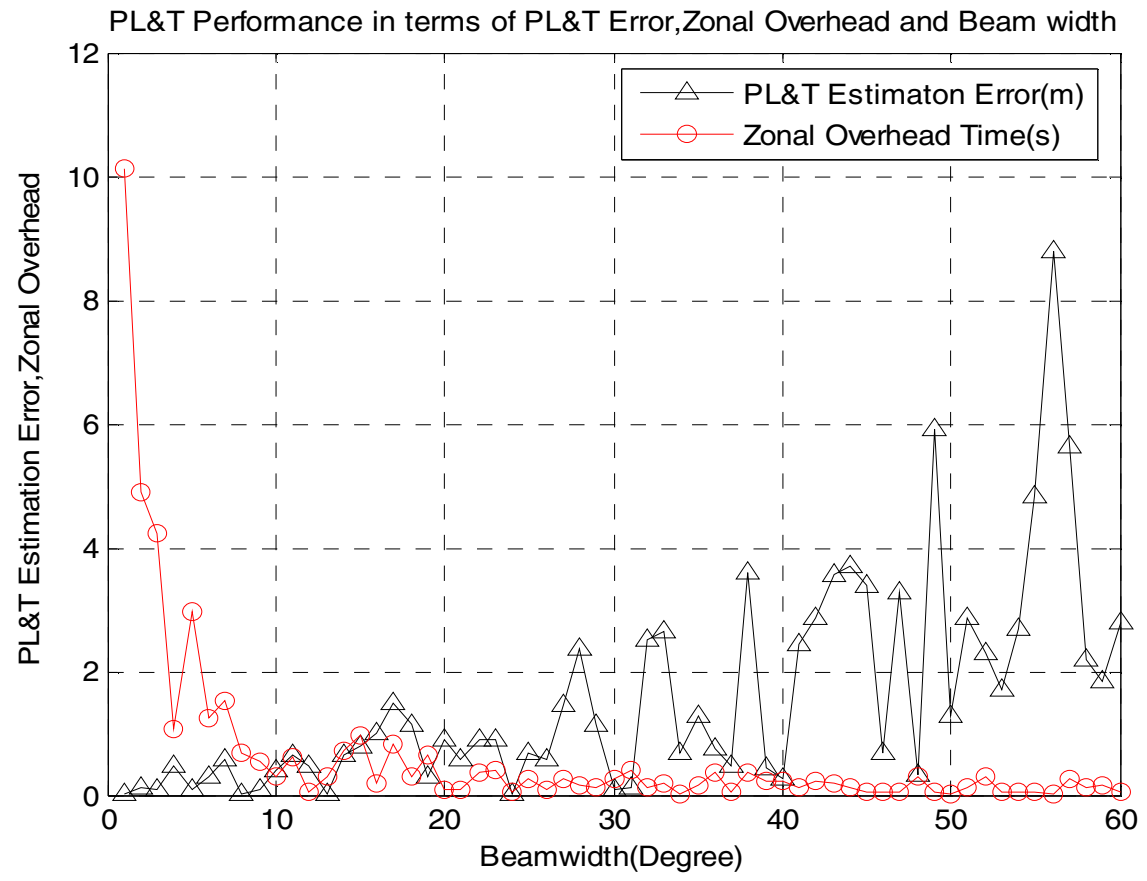

Figure 10. PL&T performance regarding Beam width

## VII. Dynamic PL&T Using Kv Transform Coding For Error Correction And Sample Interleaving

Since multi-path fading forces the tracking operation to be conducted at low Eb/No, it is essential to maintain a bounded BER with an average sustained data transmission rate. Using KV transform, it is possible to interleave discrete samples and allow one out of four discrete samples to be corrected exactly [9], it is expected that the tracking accuracy improves significantly. Figure 11-12 illustrates the tracking accuracy improvement when using KV transform that allows sample interleaving over that of without interleaving.

Dynamic PL&T has found higher tracking accuracy for bits interleaving KV during data transmission and reception against the multipath interference. Figure 11 - 12 show that using Interleaving KV transform coding, it is possible to achieve the significant tracking accuracy of over 90% using 10 db of $E_b/N_o$ for BER above $10^{-6}$ and therefore allows the dynamic PL&T algorithm to be very robust under heavily multipath faded channels with Doppler effect.

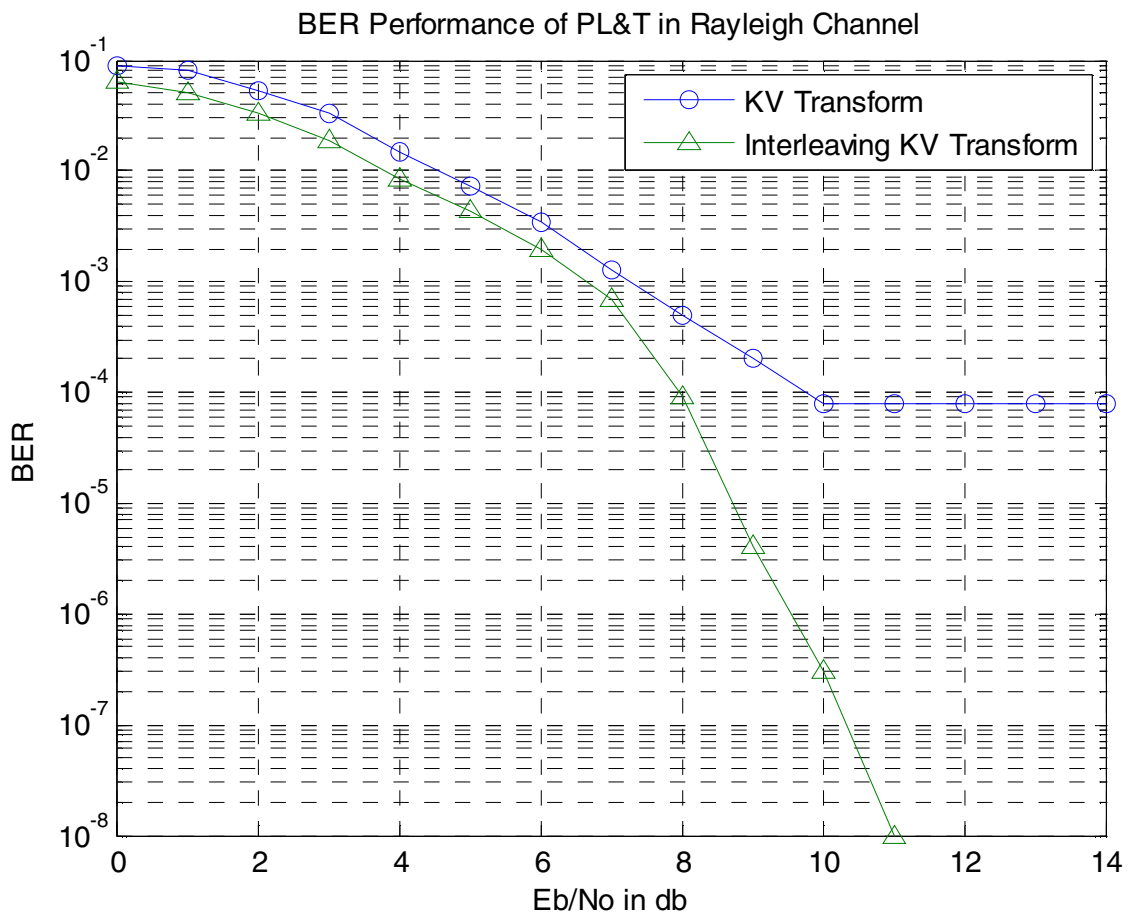

Figure 11. Improved Bit Error Rate using KV Transform

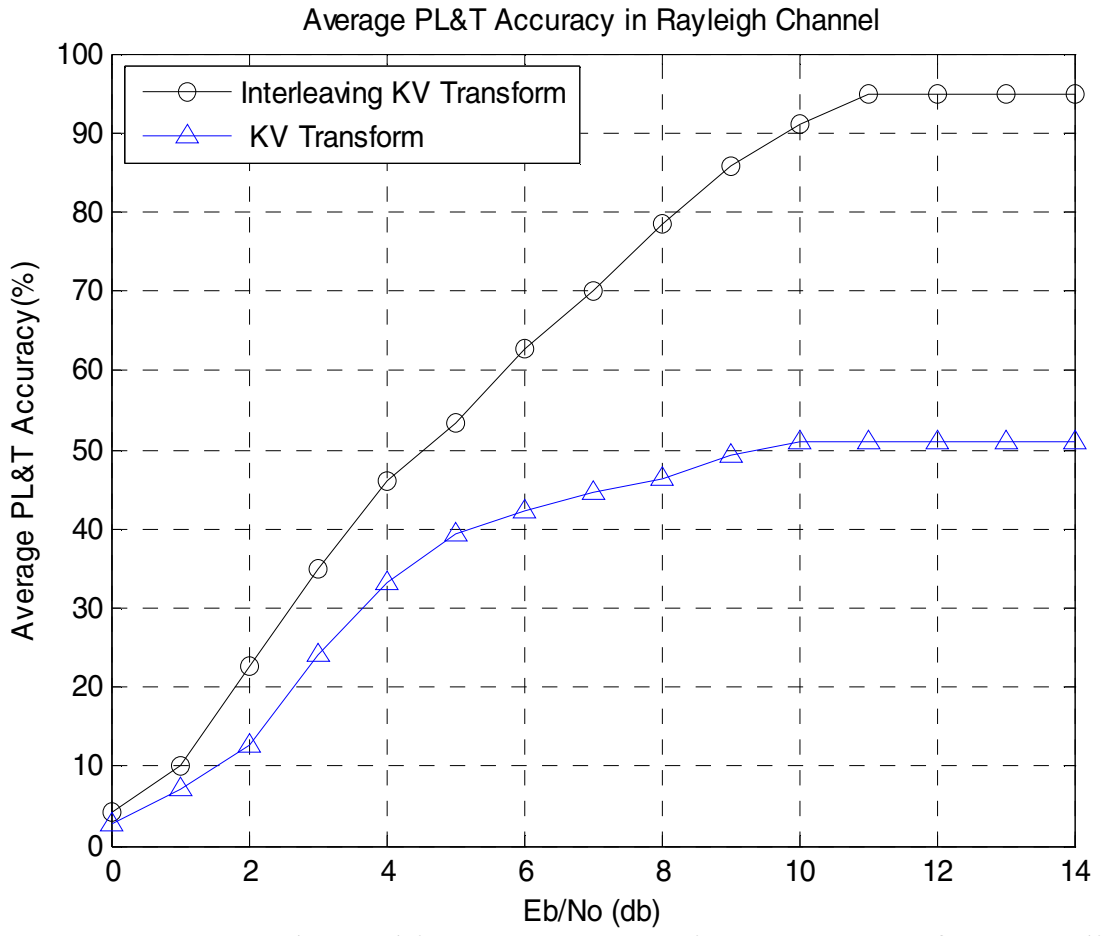

Figure 12. Improved Tracking Accuracy using KV Transform Coding

## VIII. CONCLUSION

This paper depict explicit dynamic PL&T technique using target's zone finding, adaptive beam forming over the zone and then allow the triangulation to be achieved using two reference nodes. The average PL&T error is found significantly lower in two reference nodes equipped with DA than other PL&T scheme deploying single node using DA, two nodes using OA and three nodes using OA. In addition, dynamic PL&T method has outstanding PL&T accuracy as compared to other PL&T techniques. In addition, the optimum beam width for dynamic PL&T is found 10 degree to optimize the estimation error and zonal overhead. Furthermore, the performance in terms of tracking accuracy is significantly improved in multipath faded channels when using KV transform coding and achieves greater than 90% accuracy with BERs above $10^{-6}$.

## ACKNOWLEDGEMENT

This work is supported in part by the US Army Research Office funding under Research Cooperative Agreement grant No. W911NF-10-1-0087 and NSF RISE funding un-der Contract C09-00960 to the ARO Center for Battlefield Communications (CeBCom), Department of ECE, Prairie View A&M University. The views and conclusions con-tained in this document are those of the authors and should not be interpreted as representing the official policies, either expressed or implied, of the NSF or the U. S. Government.